\documentclass[conference]{IEEEtran}
\IEEEoverridecommandlockouts
\usepackage{cite}
\usepackage{amsmath,amssymb,amsfonts}
\usepackage{graphicx}
\usepackage{textcomp}
\usepackage{xcolor}
\usepackage{subcaption}
\usepackage{multirow}
\usepackage{booktabs}
\usepackage[export]{adjustbox}

\usepackage{amssymb}
\usepackage{amsfonts}
\usepackage[linesnumbered,ruled,vlined]{algorithm2e}

\SetCommentSty{mycommfont}
\usepackage{siunitx}

\renewcommand{\baselinestretch}{.9825}    

\IEEEoverridecommandlockouts
\IEEEpubid{\makebox[\columnwidth]{978-1-6654-3922-0/21/\$31.00 $\copyright$2021 IEEE \hfill} \hspace{\columnsep}\makebox[\columnwidth]{ }}

\begin{document}

\bstctlcite{IEEEexample:BSTcontrol} 


\title{Neural Contextual Bandits Based Dynamic Sensor Selection for Low-Power Body-Area
Networks}

\author{\IEEEauthorblockN{Berken Utku Demirel, Luke Chen,  Mohammad Abdullah Al Faruque}
\IEEEauthorblockA{\textit{Department of Electrical Engineering and Computer Science} \\
\textit{University of California, Irvine, California, USA}\\ 
\textit{\{bdemirel, panwangc,  alfaruqu\}}@uci.edu}\vspace{-9truemm}}

\maketitle

\begin{abstract}

Providing health monitoring devices with machine intelligence is important for enabling automatic mobile healthcare applications. However, this brings additional challenges due to the resource scarcity of these devices. This work introduces a neural contextual bandits based dynamic sensor selection methodology for high-performance and resource-efficient body-area networks to realize next generation mobile health monitoring devices. The methodology utilizes contextual bandits to select the most informative sensor combinations during runtime and ignore redundant data for decreasing transmission and computing power in a body area network (BAN). The proposed method has been validated using one of the most common health monitoring applications: cardiac activity monitoring.
Solutions from our proposed method are compared against those from related works in terms of classification performance and energy while considering the communication energy consumption. Our final solutions could reach $78.8\%$ AU-PRC on the PTB-XL ECG dataset for cardiac abnormality detection while decreasing the overall energy consumption and computational energy by $3.7 \times$ and $4.3 \times$, respectively. 

\end{abstract}


\section{Introduction}
Machine Learning (ML) based solutions for mobile health applications have shown promising results in providing efficient monitoring of numerous physiological signals such as ECG \cite{Modema}, EEG or PPG. The increased use of multi-modal devices have led to the introduction of body area networks in wearable devices for continuous real-time monitoring. Traditionally, the main operating principle of these mobile health monitoring devices starts by sending the collected raw data from multiple sensors to an intermediate instrument (e.g., smartphone) using a low-power wireless technology (e.g., Bluetooth). Then, the data is forwarded further to a cloud server for storing or processing heavy algorithms such as Deep Neural Networks (DNN) using long-range energy-hungry wireless methods \cite{Wireless} such as Wi-Fi or LTE. However, the additional energy associated with raw data transmission over the wireless medium brings concerns for extra energy consumption of sensor devices for this approach.
Moreover, at the end of this transmission, since the redundant data is also sent to the mobile devices, the amount of data that the device is required to process increases, which decreases the device's battery life and makes it challenging to sustain continuous monitoring over long periods. As a result, the current growing trend encourages the adoption of edge computing, where most of the processing is leveraged to the closest devices to data generation. In this setting, the recording device (e.g., smart sensors) can contain a lightweight model to perform the needed processing to decide to transmit the useful or most informative part of the data.
\\
Different methods exist in the literature regarding filtering out the unnecessary transmission and processing of raw Electrocardiogram (ECG) data to wearable devices. For example, a recent study from Demirel et al. \cite{Demirel_IoT} has proposed to use early layers of the neural network to decide whether the current heart activity is normal or abnormal in terms of the beat waveform morphology and heart rate variability; if it is normal, the data is not transmitted to a cloud server for further investigation. However, this approach has not considered the multi-channel element of the ECG signals and focused on identifying small numbers of cardiac abnormalities that do not represent the complexity and difficulty of heart monitoring applications. More recent works \cite{Royal_Society, CinC_2021} have focused on finding the optimal lead subset of the 12-lead ECG at design time to eliminate the redundancy, which can help improve the generalizability of Deep Learning (DL) based models and decrease the energy consumption due to communication and classification. However, these proposed methods do not consider dynamic changes in the human heart while designing their method, which results in a static system. For example, the study \cite{Royal_Society} investigates what channels are necessary to keep and which ones may be ignored when considering an automated detection system for cardiac ECG abnormalities, and found a 4-lead ECG subset for classification of ECG abnormalities using the DL model. However, in this paper, we showed that 2-lead ECG is also enough for some signal segments and using 4-lead ECG would increase the power consumption without increasing the classification performance.
Although, the existing literature on ECG abnormality detection is extensive and focuses particularly on the multi-lead systems.

\begin{figure}[h]
\begin{center}
{\includegraphics[,width = 0.5\textwidth]{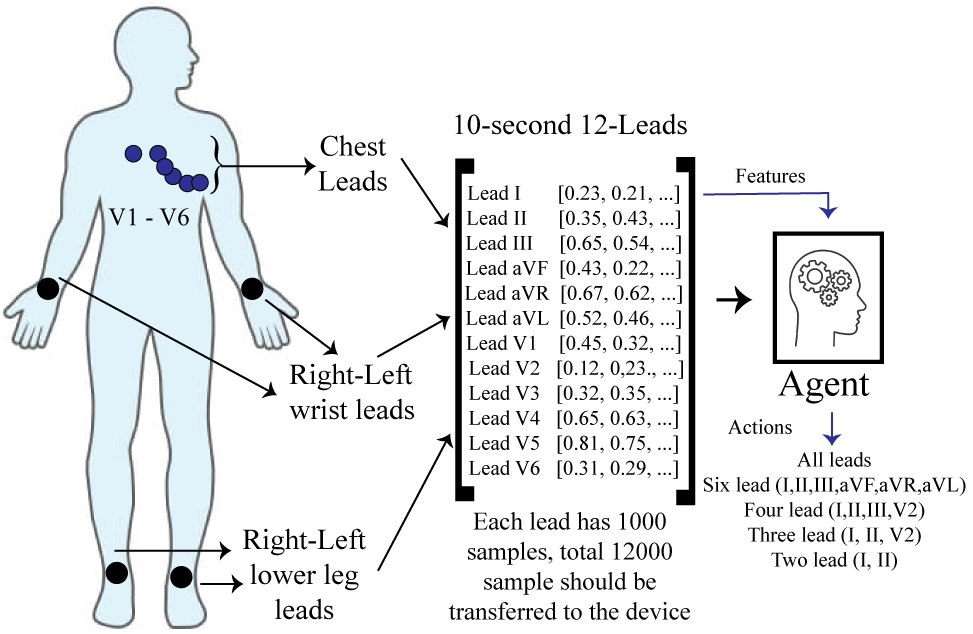}}
\end{center}
\vspace{-2ex}
\caption{The overview of proposed method} 
\label{obj}
\end{figure}

In this work, we have proposed a dynamic lead selection methodology using neural contextual bandits with thompson sampling for high-performance and low-power body-area networks.

\subsection{Motivational Case Study}

\begin{figure}[h]
    \centering
    \includegraphics[scale=0.7]{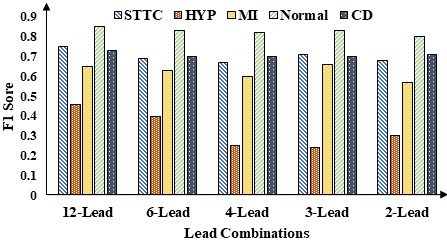}
    \caption{The classification performance of each class with different sensor combinations}
    \label{fig:motivational}
\end{figure}

We have done several experiments to show why dynamic lead selection is needed and can help decrease energy consumption significantly while classifying ECG signals with high performance in resource-constrained devices. Firstly, we examine five different classifiers, one for each lead subset (from twelve-lead, six-lead, four-lead, three-lead, and two-lead ECG recordings), to detect normal ECG (NORM), myocardial infarction (MI), conduction disturbance (CD), ST/T-changes (STTC), hypertrophy (HYP) in PTB-XL dataset \cite{wagner_ptb_xl_2020}. The lead subsets are chosen as twelve-lead, six-lead (I, II, III, aVR, aVL, aVF), four-lead (I, II, III, V2), three-lead (I, II, V2) and two-lead (I and II) ECG recordings.
While doing the experiments, we followed the suggested evaluation method for the PTB-XL dataset; training on the first eight folds and the ninth and tenth fold is used as validation and test sets, respectively. For each subset and class, the F1 score ($2 \times \frac{Precision \times Recall}{Precision + Recall}$) in percentage is obtained and reported in Figure \ref{fig:motivational}.
\vspace{-4mm}
\vspace{-2mm}
The most interesting aspect of this Figure is that the classification performance of the neural network stays relatively constant, around $80-81\%$, for normal ECG (NORM) in each lead combination. However, HYP type segments classification rate decreases heavily in 6 and 3 lead combinations, especially 3-lead is the lowest classification rate with $20\%$ drop compared to the single lead. Although it looks like STTC and CD detection performance stays close to constant in different lead combinations, its percentage is decreased by $7\%$ from the best one in a single lead. Surprisingly, these results show no direct relation with increasing the number of sensors for the classification of different diseases. Also, it should be noted that the classifier designed with less number of sensors, such as 2-lead, has $6\times$ fewer parameters and Floating point operation (FLOPs) compared to the baseline model that uses all available sensors. Moreover, using all available sensors brings additional energy consumption due to transferring raw data to the device for processing. As communication accounts for a significant portion
of the total power consumption of a connected device \cite{ISLPED_3}, a method to decrease the communication data will enhance battery
life significantly and realize the ultra-low power Body-Area Networks.
Overall, these results suggest that if the sensor combinations are selected precisely during runtime, significant energy consumption can be prevented without sacrificing the classification performance.

\subsection{Problem Statement and Research Contributions}

Based on the previous arguments, the research challenges we aim to address in this work include:
\begin{itemize}
    \item What techniques should be utilized to choose the optimal lead subset using the current condition of the data during runtime?
    \item How to design the dynamic lead controller, which should be energy and memory efficient while executing the required task with low latency to provide real-time performance?   
\end{itemize}
Addressing the challenges mentioned above, we propose a contextual bandit based dynamic lead selection method that chooses the most informative or necessary channels for the current ECG signal and ignores the redundant raw data before transmission. While designing the agent that selects optimal leads, we have considered multiple design objectives, which are accuracy and energy consumption in communication and classification. 
Our research contributions can be summarized as follows:
\begin{itemize}
    \item We proposed a dynamic lead selection methodology using neural contextual bandit for high-performance and resource-efficient mobile health applications.
    \item We proposed a data driven novel contextual bandits environment that uses a simple feature extracted from different sensors in ECG signals to determine which lead combinations can be chosen to decrease communication and computational overhead without sacrificing the classification performance.
    \item We demonstrate the effectiveness of our proposed methodology in a body area network to monitor cardiac activity.
    \item On the PTB-XL ECG dataset \cite{wagner_ptb_xl_2020}, our proposed methodology achieves state-of-the-art performance metrics for classifying cardiac activity into five diagnoses, reaching $78\%$ AU-PRC while decreasing the overall energy comsumption and computational energy by $3.7 \times$ and $4.3 \times$ respectively . 
\end{itemize}

\vspace{-3mm}
\section{Preliminaries and Related work}

\subsection{Contextual Bandits}
The contextual bandit is a sequential decision-making problem in which, at every time step, the learner observes a context, chooses one of the possible actions (arms), and receives a reward for the chosen action. This problem has been extensively studied, where two prominent solutions namely Upper Confidence Bound (UCB) and Thompson Sampling (TS) have been adapted to a wide range of applications, from e-commerce and recommender systems \cite{Bandits} to medical trials. More recently, with the advancements in deep learning, researchers started applying neural networks to the contextual bandits problem. Neural versions of UCB and Thompson sampling have been proposed with theoretical guarantees on performance \cite{NeuralUCB, NeuralTS}. In this work, we applied the Neural TS (Thompson sampling) algorithm, or NTS for short, from \cite{NeuralTS} as it is more efficient compared to its UCB counterpart, which is desirable for a real-time problem context. The goal of the contextual bandit decision-maker is to minimize the difference in total expected reward collected when compared to an optimal policy and a quantity termed regret. For example, consider an advertisement engine in an online shopping store, where the context can be the user's query, the arms can be the set of millions of sponsored products, and the reward can be a click or a purchase. In our problem formulation of contextual bandits, the different sensor combinations are represented as arms, and the contexts are defined as extracted features from ECG segments.



\subsection{Cardiac Activity Monitoring}
Cardiovascular disease (CVD) is one of the leading causes of mortality worldwide. In addition, according to the National Center for Health Statistics \cite{N0_2}, the age-adjusted death rate of CVDs was 219.4 per 100,000 in 2017, which equalled 859,125 dead and 2.2 million people hospitalized in the United States. It is now well established from various studies that the death toll due to CVDs is more than cancer and chronic lung disease combined \cite{N0_4}. There is a growing body of literature that recognizes the importance of ECG based cardiac activity monitoring \cite{ISLPED_2} in real-time. Although some works investigate which sensor combinations can be helpful and enough for the arrhythmia classification, there is no work to select leads dynamically in real-time to reduce the system's power consumption in terms of communication and computation.

\section{Our Proposed Methodology}

\subsection{Adaptive CNN Classifier}
\begin{figure}[h]
    \centering
    \includegraphics{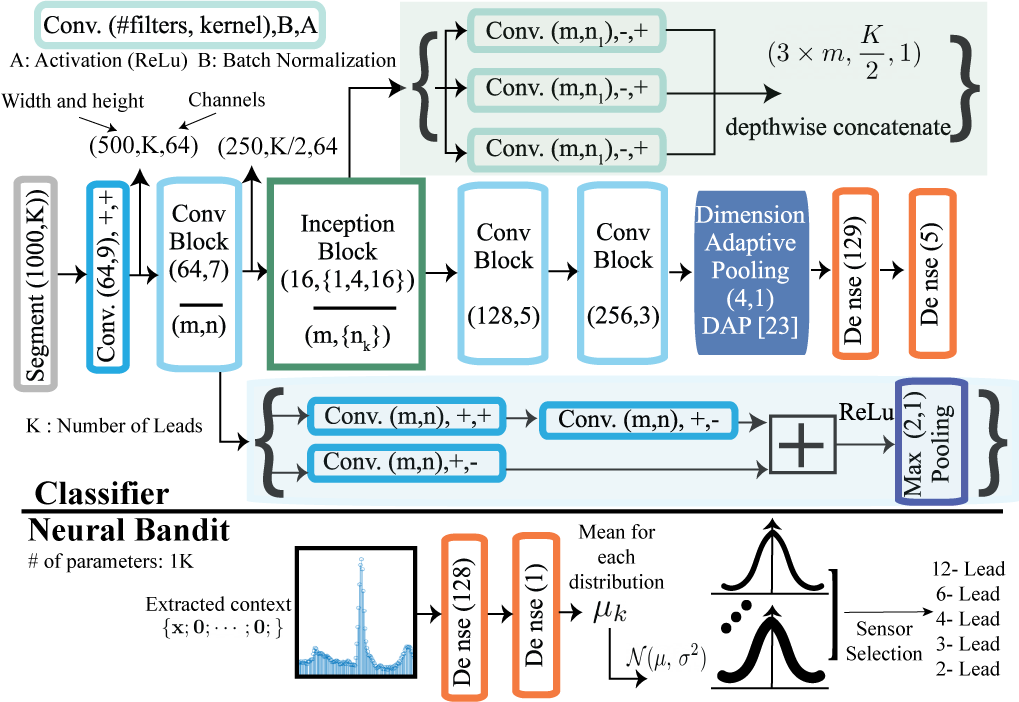}
    \caption{Proposed Classifier and Neural Bandit Architecture}
    \label{fig:arch}
\end{figure}

The designed classifier is a convolutional neural network, which take as input only the raw 10-second ECG segments and no other patient- or ECG-related features, and classifies a single segment into five classes (NORM, MI, CD, STTC, HYP). The architecture is designed to extract various morphological features from the complete segment by employing different lengths of convolutional filters at different layers of the architecture. While designing the architecture, we utilized both the Residual connections introduced by He et al. in \cite{ResNet} and the Inception architecture \cite{Inception} that has been shown to achieve good performance while maintaining computational and memory costs at low levels.
The Conv blocks in Figure \ref{fig:arch} show the implemented original residual connections where the activation is applied after addition. The model includes three residual blocks with different kernel sizes and filter numbers. Every residual block subsamples its inputs by a factor of 2 by taking the maximum sample (i.e., max pooling with stride 2). The Rectified Linear Unit (ReLu) is utilized as the activation function in the classifier. All convolutional layers are implemented using a stride of 1 except the first filter, which moves two samples, resulting in half of the samples $K/2$. Unlike the Inception architecture, the wider layer (one after the first residual block) is not stacked up together; instead, we have used Residual blocks, which helps to reduce the dimension of the network while combining the various features of a heartbeat.

\vspace{-3mm}
Otherwise, the sequential connections of these wider layers result in a quadratic increase of computation and parameters, making the network inefficient and prone to overfitting.
\\
As the number of channels is controlled during runtime, the input to classifier which is represented as $(1000,K,1)$, can be different. The K value in this representation is the number of sensors which the neural bandits choose dynamically during runtime. However, CNN architectures are generally designed to work with incoming data of a fixed channel size, and changes in the sizes of their inputs cause substantial performance loss, unnecessary computations, or failure in operation. To handle these limitations, we have utilized the dimension-adaptive pooling (DAP) layer \cite{DANA} that makes DNNs flexible and more robust to changes in sampling rate. After each convolutional layer, we applied batch normalization \cite{batch} and/or a rectified linear activation. The '$+$' and '$-$' signs near the convolutional layer shows whether these operations are applied for that convolutional operation. We also used Dropout \cite{Dropout} before the last dense layer with a probability of $0.5$ to prevent overfitting. The final fully connected sigmoid layer produces a distribution over the five output classes.
\vspace{-2mm}
\subsection{Neural Thompson Sampling}
To apply NTS, we consider the problem of dynamic lead selection as a contextual $K$-armed bandit problem, where each arm represents a combination of leads and we have a finite number of rounds $T$. At every round $t\in{[T]}$, the agent observes $K$ contextual vectors of size $d$ $\{\mathbf{x}_{t,k}\in{\mathbb{R}^d}|k\in{K}\}$. When the agent selects an arm $a_t$ it receives a corresponding reward $r_{t,a_t}$. The goal is to maximize the total expected reward or in other words to minimize the sum of regrets:

\begin{align}\label{eq:regret}
    R_T = \mathbb{E}[\sum_{t=1}^T(r_{t,a_t^*} - r_{t,a_t})]
\end{align}

where $a_t^*$ is the optimal arm at round $t$ which gives the maximum expected reward.

Thompson sampling works by associating each arm with a reward distribution, for simplicity, we use the Gaussian distribution in accordance with \cite{NeuralTS}. The parameters of distributions are defined through its mean and variance where the trained neural network approximates the mean for each arm. And the variance is approximated using the approximated mean with Equation \ref{eq:variance}. The network is updated every time a reward is observed through gradient descent on the squared loss between the networks predicted reward and the real reward.

\begin{align}\label{eq:variance}
    \sigma_{t,k}^2 = \lambda \mathbf{g}^T(\mathbf{x}_{t,k};\mathbf{\theta}_{t-1})\mathbf{U}_{t-1}^{-1}\mathbf{g}(\mathbf{x}_{t,k};\mathbf{\theta}_{t-1})/m \\
    \mathbf{U}_t = \mathbf{U}_{t-1} + \mathbf{g}(\mathbf{x}_{t,a_t};\mathbf{\theta}_t)\mathbf{g}(\mathbf{x}_{t,a_t};\mathbf{\theta}_t)^T/m
\end{align}

Where $\lambda$ is the regularization parameter, $\mathbf{g}$ is the gradient list, $\mathbf{U}$ is the neural tangent kernel and $m$ is the width of the neural network.

\subsection{Context Extraction}
While extracting the contexts, we have investigated different features of ECG signals. At the end, it is found that the trained neural bandits performed the best when the contexts are extracted from a signal segment. For extracting the beat from 10-second 12-lead signals, we have only used lead-II. Although numerous robust methods have already been available for R peak detection, we have used the Pan-Tompkins algorithm [29], a real-time QRS complex-based heartbeat detection approach that has an accuracy of up
to 99.5\%, and the algorithm for improving R-peak detection is beyond the scope of this manuscript. Similar to other work \cite{Demirel_IoT}, we take $0.25$ second before the peak and $0.3$ second after the peak (totally $0.55$ second containing the R peak) are used to represent the corresponding heartbeat. As the last step for preprocessing, the segmented beats are normalized to have a maximum value of 1 before feeding them to the neural bandit.

\section{Experimental Setup}

\subsection{Training Classifier}\label{train_classifier}
We use data from the PTB-XL dataset \cite{wagner_ptb_xl_2020} which comprises 21837 clinical 12-lead ECG records of 10 seconds from 18885 patients, where 52\% were male and 48\% were female. The ECG statements were assigned to three non-mutually exclusive categories diag (short for diagnostic statements such as "anterior myocardial infarction"), form (related to notable changes of particular segments within the ECG such as ``abnormal QRS complex") and rhythm (related to certain changes of the rhythm such as "atrial fibrillation"). In total, there are 71 different statements, which decompose into 44 diagnostic, 12 rhythm and 19 form statements, 4 of which are also used as diagnostic ECG statements. A hierarchical organization into five coarse superclasses and 24 sub-classes is also provided for diagnostic statements. The interested readers can be referred to the original publication for further details on the dataset, the annotation scheme, and other ECG datasets. In summary, PTB-XL stands out by its size as the to-date largest publicly accessible clinical ECG dataset and through its rich set of ECG annotations and other metadata, which turns the dataset into an ideal resource for the training and evaluation of machine learning algorithms. Throughout this paper, we use the recommended train-test splits provided by PTB-XL \cite{wagner_ptb_xl_2020}, which is training on the first eight folds and the ninth, tenth fold is used as validation and test sets, respectively. Also, the input data is used at a sampling frequency of 100 Hz.
\\
Moreover, since the size of each input is controlled during runtime, the CNN classifier should adapt to the changes in the sampling rate. Therefore, we have used adaptive dimension training, which comprises dimension randomization and optimization with accumulated gradients as in \cite{DANA}. This process works by training the CNN on input data of several randomly selected dimensions (different lead combinations). In this way, the model can learn the waveform morphological features of different sizes of heartbeats. 
\\
The classifier network is trained with Glorot initialization of the weights \cite{glorot}. L2-regularization with $0.0002$ is applied for each convolution operation for Inception while the last linear layers are trained with a $0.00005$ L2 value. We used the Adam optimizer \cite{Adam} with the default parameters $\beta_1=0.9$ and $\beta_2=0.999$, and a mini-batch size of 80. The learning rate is initialized to $0.001$.

\subsection{Training Bandits}

We define the optimal action $a_t^*$, given a context $x_t$, as the one which produces the correct classification result and uses the least amount of leads. In the case that none of the lead combinations can classify correctly, we choose the option with the least amount of leads, in other words the lowest energy consumption. However, it is observed that if the same training data which is employed before for training the ML model is also used for the contextual bandits, the agent becomes highly biased since the optimal arm for each round depends on the correct classification of the current ECG segment and the trained model has already seen those data before in the training phase. This biasing leads to an action space that is highly concentrated in a single action. To prevent this biasing, we construct a mixed training dataset by combining the training and validation where we ignore the majority action samples in the training folds and keep the validation dataset as it is. At the end, it is observed that the combined dataset decreases the biasing in the neural bandits. The training, validation, and test data for the bandit agent follows the same scheme from \cite{wagner_ptb_xl_2020} as mentioned previously in Section \ref{train_classifier}.

The contextual bandit neural network is a single hidden layer network with input size equal to $K$ arms times $1000$ features, 128 hidden units, and 1 output node to approximate the mean reward. The input contextual vector is constructed following \cite{NeuralTS} such that each arm sees a corresponding vector in the form $\mathbf{\{x; 0;\cdots; 0;\}}$ for arm 0, $\mathbf{\{0; x;\cdots; 0;\}}$ for arm 1, up to $K$ where $x$ is the feature vector. Two parameters used in NTS are $\nu$ and $\lambda$ for which we set to 1e\textsuperscript{-6} and 1e\textsuperscript{-1} based on a grid search over $\{$1, 1e\textsuperscript{-1}, 1e\textsuperscript{-2}, 1e\textsuperscript{-3}$\}$ for $\lambda$ and $\{$1e\textsuperscript{-1}, 1e\textsuperscript{-2}, 1e\textsuperscript{-3}, 1e\textsuperscript{-4}, 1e\textsuperscript{-5}, 1e\textsuperscript{-6}$\}$ for $\nu$. Each context is trained for 100 iterations of stochastic gradient descent with learning rate 1e\textsuperscript{-2} and weight decay $\lambda/counter$ where the counter increases by 1 up until the total training samples.

\section{Experiments and Results}
To evaluate the classification performance, we have used the macroaverage precision, macro-average recall, accuracy, and Area under the Precision-Recall operating characteristic curve (AU-PRC) metrics. The evaluation results of our adaptive performance on PTB-XL dataset is given in Table 2. As shown in the Table, we may conclude that the overall performance of our proposed method overperforms or reaches the baseline model in all leads for ECG classification tasks while being energy and memory efficient.

\begin{table}[h]
\centering
\caption{Performance Comparison of Proposed method with Related Works}
\vspace{3mm}
\begin{adjustbox}{width=\columnwidth,center}
\label{tab:performance}
\renewcommand{\arraystretch}{1.5}
\begin{tabular}{lcccc}\toprule
\textbf{Work}  & \textbf{Lead} & \textbf{Precision} & \textbf{Recall} & \textbf{AU-PRC} \\ 
\hline
\textbf{ML-ResNet \cite{ML-ResNet}}  & 12-Lead & 66 & \textbf{75} & 77.9  \\ 
\textbf{MFB-CBRNN \cite{CBRNN}}   & 12-Lead & 65 & 68 & 78.6 \\
\textbf{FCNN \cite{FCNN} } & 12-Lead & 68.1 & 70.5 & 77.4 \\
\textbf{A-CNN \cite{A-CNN}}  & 12-Lead & 65.8 & 68.9 & 77.6 \\
\textbf{Ours}  & 2-Lead & 68 & 61 & 71.1 \\
\textbf{Ours}  & 12-Lead & 71 & 66 & 78.1 \\
\textbf{Ours} & Adaptive & \textbf{72} & 63 & \textbf{78.8} \\ 
\bottomrule
\end{tabular}
\end{adjustbox}
\end{table}

Especially for abnormality detection task, the baseline classifier which uses all available sensors (12-leads) performs worse than our proposed adaptive method while increasing the computational complexity of the model more than $6.4 \times$. This shows that using all sensors brings additional energy consumption due to classification and communication while decreasing the performance.  Moreover, when the proposed method is compared with the related works, it is noticeable that it outperforms or reaches the performance while decreasing the overall calculation for classifiers.
\\
Although the adaptive sensor selection decreases the overall computation and communication while achieving comparable classification performance, it should be emphasized that one of the more significant contributions to emerge from this study is that our proposed method is not a substitute for other methods concerning resource-constrained low-power devices; instead, it is a complementary method that can be used along with them. For example,  hardware-aware hyper-parameter
optimization \cite{hyper} or pruning and quantization \cite{HW2} of deep learning models are widely used in literature, once those dynamic compression techniques have compressed a network, our adaptive sampling can still be applied to the multi-modal systems and fed to the compressed network. Or, different deep learning architectures which are concerned about energy and memory are recently proposed for ECG classification, our proposed method can be utilized with these architectures as an additional preprocessing step.

\section{Memory and Energy Consumption Evaluation}

We evaluate the proposed method's memory footprint and energy consumption on the EFM32 Giant Gecko ARM Cortex-M3-based 32-bit microcontrollers (MCUs), which has a 1024 kB flash and 128 kB of RAM with CPU speeds of up to 48 $MHz$.
Table~\ref{tab:Hardware} shows the execution time, energy consumption, and required memory for each operation that runs on the edge device which is deployed to the body area network for selecting sensors during runtime. The operations are implemented and deployed to the target device using MATLAB (MATLAB and Coder Toolbox Release R2021b, The MathWorks, Inc, USA).

\begin{table}[h]
\caption{Memory Footprint, Execution Time and Energy Consumption Evaluation on EFM32 Giant Gecko Development Board.}
\begin{adjustbox}{width=\columnwidth,center}
\begin{tabular}{cccccc}
\toprule
Operations   & \begin{tabular}[c]{@{}c@{}}Exe.\\ Time (ms)\end{tabular} & \begin{tabular}[c]{@{}c@{}}Avg.\\ Energy ($\mu$J)\end{tabular} & \begin{tabular}[c]{@{}c@{}}Flash Memory \\ Footprint (KB)\end{tabular} & \multicolumn{1}{c}{\begin{tabular}[c]{@{}c@{}}RAM Memory\\ Footprint (KB)\end{tabular}} \\ \hline
Context extraction & 1200  & 580.2 & 9.4 & 20.7 \\ 
Post-processing & 2.7  & 1.17 & 8 & 43 \\ 
Neural Bandit & 175  & 8  & 32.5  & 35.7 \\\hline
Overall & 1377.7 & 589  & $\geq$ 64 KB  & $\geq$ 64 KB \\\hline \hline
\end{tabular}
\end{adjustbox}
\label{tab:Hardware}
\end{table}

\begin{figure}[h]
    \centering
    \includegraphics[scale=0.68]{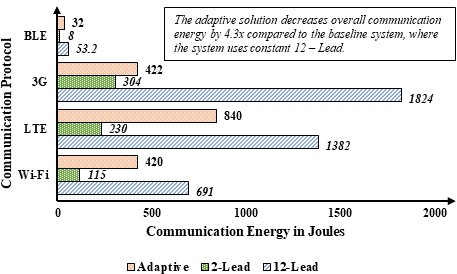}
    \caption{The energy consumption of transfer operation for different lead combinations and proposed adaptive solution}
    \label{fig:Comm_energy}
\end{figure}

The overall execution time for a $10$-second ECG segment, which is sampled from the PTB-XL dataset takes $1377.7$ ms in the edge device with $589$ $\mu$J energy consumption. As the context extraction operation includes filtering, peak-detection, segmentation and normalization, its computational overhead dominates the overall operations. Also, our proposed method is compatible with any edge device with a minimum RAM of $64$ KB. As a result, our method guarantees high performance while maintaining the low-power edge devices' requirements of being resource-efficient in terms of energy and memory. Moreover, We have done several experiments to show the advantages of the proposed method for decreasing the communication energy of the body area networks. The 1-hour raw 12-lead, 2-lead and dynamically selected lead combinations are transferred to the mobile devices without changing the sampling rate of the signal to mimic the working principle of the body area networks. Then, the energy consumption of these transfer operations is calculated for four different communication technologies, \textit{Wi-Fi}, \textit{LTE}, \textit{3G} and \textit{BLE}. The energy consumption for all cases' transfer operation is shown in Figure \ref{fig:Comm_energy}.

While calculating the energy consumption of these cases, we have followed the $\mu J/bit$ values given in \cite{Demirel_IoT} for the \textit{Wi-Fi}, \textit{LTE}, and \textit{3G}. For \textit{BLE} protocol energy consumption, we performed the profiling for data exchange on an EFR32BG13 Blue Gecko Bluetooth® Low Energy SoC which has 32-bit ARM Cortex-M4 core with 40 MHz maximum operating frequency. What stands out in Figure \ref{fig:Comm_energy} is that when the proposed adaptive method is utilized instead of using common 12-lead sensor combination in the body area network, the communication energy savings can reach $4.3 \times$ in all protocols. Although, the energy consumption of the 2-lead raw data is the lowest, its classification cannot achieve a comparable performance with state-of-the-art works.
While comparing the FLOPs, we have only considered the architectures which are given explicitly. Furthermore, any preprocessing such as Fourier or wavelet transforms are ignored, and only the CNN architecture for classification FLOPs is calculated. 

\begin{figure}[h]
    \centering
    \includegraphics[scale=0.63]{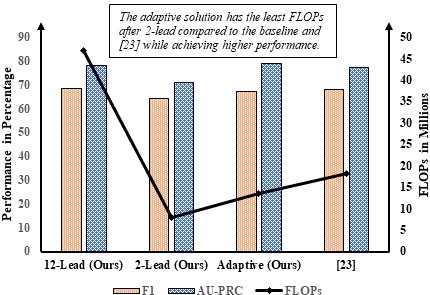}
    \caption{Comparison of computational complexity and
performance}
    \label{fig:FLOPs_perf}
\end{figure}
\vspace{-2mm}

While calculating the total number of floating-point operations (FLOPs), we have followed \cite{NVDIA} where the convolution is assumed to be implemented
as a sliding window and that the nonlinearity function is computed for free. For convolutional kernels we have;
\vspace{-2mm}
\begin{equation}
    \text{FLOPs} = 2HW(C_{in}K+1)C_{out} 
\end{equation}
Where $H$, $W$ and $C_{in}$ are the height, width and number of channels of the input feature map, $K$ is the kernel width, and $C_{out}$ is the number of output channels. As shown in Figure \ref{fig:FLOPs_perf}, the adaptive solution reduces the overall FLOPs value by 1.35$\times$, and 3.5$\times$ compared to \cite{FCNN} and baseline where all available sensors are used. Moreover, while decreasing overall computation, the adaptive solution does not sacrifice any performance even the overall F1 score value increases 10$\%$.
\vspace{-4mm}
\section{Conclusion}

In this paper, we propose a methodology for dynamic sensor
selection of body area networks in low-power resource-constrained devices in terms of memory and battery (e.g. wearable devices) using neural contextual bandits. Moreover, we also presented a data driven novel multi armed bandits that uses a simple feature extracted from different sensors in ECG signals to determine which lead combinations can be chosen to decrease communication and computational overhead without sacrificing the classification performance. Evaluation on the PTB-XL dataset shows that our proposed dynamic sensor selection solution requires 64 KB of RAM and achieves up to $3.7 \times$ and $4.3 \times$  overall energy efficiency in computational and communication energy, respectively without sacrificing any classification performance.

\bibliographystyle{IEEEtran}
\scriptsize{
    \bibliography{sample.bib}
}

\end{document}